\documentstyle[12pt]{article}
\def\beqar {\begin{eqnarray}}
\def\eeqar {\end{eqnarray}}
\def\beq {\begin{equation}}
\def\eeq {\end{equation}}
\def \ep {{\epsilon}}
\def \half {{\textstyle{1\over 2}}}

\def \la {{\langle}}
\def \ra {{\rangle}}

\def \Tr {{\rm Tr}}

\def \S {{\cal S}}
\def \cD {{\cal D}}
\def\del{{\partial}}
\begin{document}

\begin{titlepage}
\null\vspace{-62pt}

\pagestyle{empty}
\begin{center}
\rightline{}
\rightline{CCNY-HEP-01/10}

\vspace{1.0truein} {\Large\bf Gravitational fields on a
noncommutative space}\\

\vspace{1in} V. P. NAIR\\
\vskip .1in {\it Physics Department\\ City College of the CUNY\\
New York, NY 10031}\\
\vskip .05in {\rm E-mail: vpn@ajanta.sci.ccny.cuny.edu}\\
\vspace{1.5in}

\centerline{\large\bf Abstract}
\end{center}

Noncommutative three-dimensional gravity can be described in
terms of a noncommutative Chern-Simons theory. We extend this
structure and also propose an action for gravitational fields on
an even dimensional noncommutative space. The action is worked
out in some detail for fields on a noncommutative
${\bf CP}^2$ and on $S^4$.

\end{titlepage}

%%%%%%%%%%%%%%%%%%%%%%%%%%%%%%%%%%%%%%%%%%%%%%%%%%%%%%
\pagestyle{plain}
\setcounter{page}{2}
\baselineskip =18pt
\vskip .1in\noindent{\large\bf 1. Introduction}
\vskip .1in
The dynamics of particles and fields on noncommutative spaces
have been the subject of a large number of recent publications.
Properties of gauge and other field  theories on such spaces,
their perturbative analyses, special solutions as well as the
quantum mechanics of particles have been extensively studied
\cite{general}. Field theories on some intrinsically defined
noncommutative space are certainly interesting but they become
more relevant with the realization that noncommutative spaces
arise naturally as brane configurations in the matrix model of
M-theory and also as special solutions in string theory
\cite{general,BFSS, IKKT,taylor}.  Fluctuations of branes are
generally described by gauge theories and this has been the main
motivation for the study of noncommutative field theories. In
this paper, we discuss the gravitational fields on a
noncommutative space. In three dimensions, gravity can be
described as a Chern-Simons (CS) gauge theory. Three-dimensional
gravity on a noncommutative space using noncommutative CS gauge
theory has been suggested in \cite{banados}.  We consider CS
gravity for more general odd dimensional noncommutative spaces
with the time coordinate continuous and commutative. We also
propose an action for even dimensional spaces where all
dimensions are noncommutative. This may be relevant to
noncommutative brane solutions in the matrix model for the type
IIB string theory \cite{IKKT}. The equations of motion are an
algebraic  set of equations; two special cases, namely
noncommutative ${\bf CP}^2$ and $S^4$, are considered in some
detail. We also point out an  intriguing possibility of
dynamically changing the topologies and dimensions of the base
spaces on which  the theory is defined. The noncommutative branes
which arise in the matrix model have an  underlying Lie algebra
structure and so we shall use such a framework for our
presentation. This is not a real limitation. It is actually
possible to go to a noncommutative space with the Heisenberg
algebra as the underlying structure (such as the noncommutative
plane) by taking an appropriate limit of our formulae. The
construction of gravity on noncommutative ${\bf R}^{2n}$ using
the star product formalism has been considered before
\cite{moffcham}. There is also the well known proposal by Connes
of using noncommutative geometry  for gravity \cite{Connes}.  It
is not clear at this time if our actions are related to his
proposals.
\vskip .1in
\noindent{\large\bf 2. The data for gravity}
\vskip .1in

We consider a noncommutative space with coordinates $x_i$ ,
$i=1,2,...,d$, which are realized as $(N\times N)$-matrices. The
commutators
$[x_i,x_j]\equiv \omega_{ij}$ can be expanded in terms of the Lie
algebra
$\underline{U(N)}$ of $U(N)$. We can now consider the commutators
of pairs of different $\omega_{ij}$'s or of $\omega_{ij}$ with
$x_k$, to define new elements of $\underline{U(N)}$ and continue
until we have a closed algebra. Then $(x,\omega, ...)$ will form
the $(N\times N)$-representation of a Lie algebra $\underline{G}$
which is naturally embedded in $\underline{U(N)}$. The $x_i$'s
themselves may be considered as belonging to $\underline{G}
-\underline{H}$, describing a $G/H$-space in the commuting limit.
Thus, starting with the $x_i$'s as $(N\times N)$-matrices, this
is the emerging structure that we should be considering. In the
following, we shall actually consider a more specialized case
where $G/H$ is a symmetric space, so that 
$[x_i,x_j]=-\omega_{ij}$ belong to $\underline{H}$. (If this is
not the case, we may have to consider a larger set of $x_i$ for
which this is the case and then append some algebraic conditions
on the $x_i$'s.)  The group $G$ is naturally embedded in $U(N)$,
but in general we can find a unitary group $U(k)$ such that 
$G \subseteq U(k) \subseteq U(N)$.  A typical example would be to
consider
$G=U(3)\sim SU(3)\times U(1)$ and
$H=U(2)\times U(1)$ with
$U(2)\subset SU(3)$ and
$x_i$ represented by the matrices $t_i, t_i^\dagger \in
\underline{G} -\underline{H}$. We can take them to be
$(N\times N)$-matrices in the sequence of completely symmetric
representations of $SU(3)$ with $N= \half (s+1)(s+2)$,
$s=0,1,...$.
In the commutative limit of large $N$ this corresponds to ${\bf
CP}^2$. Another example would be a noncommutative $S^4
=SO(5)/SO(4)$ described by matrices belonging to the algebra
$\underline{O(5)}-\underline{O(4)}$. In this case, $G=SO(5)
\subset U(4)$. (More details about these spaces can be found in
\cite{taylor, brane, ram}.)

Scalar functions  defined on a commutative $G/H$-space are
functions on $G$ which are invariant under the subgroup $H$.
Functions on a noncommutative space can be defined in a similar 
manner. For example, a function $f$ on the noncommutative 
$S^2$ can be represented
as $(N\times N)$-matrices, say with elements $f_{mn}$. We may think
of it as an operator ${\hat f}$ on the $N$-dimensional
representation space. One presentation of such matrices, 
useful for large $N$
considerations, can be obtained as follows. Let 
${\cal D}^{(j)}_{mk} (g)$ be the Wigner ${\cal D}$-functions
for $SU(2)$. These are the spin-$j$ matrix representation of
a group element $g =\exp(i\sigma_i \theta_i /2)$, where
$\sigma_i$ are the Pauli matrices. On the ${\cal D}$'s, 
one has the standard two sets of $SU(2)$-transformations
\beqar
J_a \cdot {\cal D}^{(j)}_{mk}(g) &=& \left[ 
{\cal D}^{(j)}_{mk}\left({\sigma_a\over 2}g\right)
\right]_{mk} \nonumber\\
K_a \cdot {\cal D}^{(j)}_{mk}(g) &=& \left[ 
{\cal D}^{(j)}_{mk}\left(g{\sigma_a\over 2}\right)
\right]_{mk} \label{a1}
\eeqar
$J_a$ will correspond to the usual angular momentum. The
action defined by the $K_a$'s is what is important for
defining a function, derivatives, vectors, etc.
In what follows, unless otherwise specified,
$\underline{G}$, $\underline{H}$ will
refer to these generators, the $K$'s.
A function on noncommutative $S^2$ may be represented as
\beq
f(g,g') = \la g \vert {\hat f}\vert g' \ra =
\sum_{mn} f_{mn} {\cal D}^{*(j)}_{mj} (g)
{\cal D}^{(j)}_{nj} (g') \label{a2}
\eeq
$N=2j+1$ and we consider fixed value of $j$.
The classical function or symbol corresponding to
${\hat f}$ is
\beq
f(g,g) = \la g \vert {\hat f}\vert g \ra =
\sum_{mn} f_{mn} {\cal D}^{*(j)}_{mj} (g)
{\cal D}^{(j)}_{nj} (g) \label{a3}
\eeq
Since the right index on the ${\cal D}$'s is fixed to
be $j$, the $K_3$-value of $f$ is zero
and this is invariant under $K_3$, which is
$\underline{H}$ in this example. $f$ may thus be considered
a function on the coset $S^2 = SU(2)/U(1)$.
One could also fix the right index to some other value
and still have $K_3 =0$. This means that there are 
different ways of
representing functions on noncommutative spaces in terms of the 
${\cal D}$-functions. We will choose (\ref{a2}).

On the noncommutative $S^2$ the product of functions is given by the matrix
product. In terms of the representation (\ref{a2}),
we can write this as
\beqar
{\hat f} {\hat h} &=& \int d\mu (g') \la g\vert {\hat f} \vert g'\ra 
\la g' \vert {\hat
h} \vert g''\ra\nonumber\\
 &=&\int d\mu (g')  \sum_{mnkl} f_{mn} h_{kl} 
{\cal D}^{*(j)}_{mj} (g){\cal D}^{(j)}_{nj} (g')
{\cal D}^{*(j)}_{kj} (g'){\cal D}^{(j)}_{lj} (g'')\nonumber\\
&=& \sum_{mnl} f_{mn} h_{nl}{\cal D}^{*(j)}_{mj} (g)
{\cal D}^{(j)}_{lj} (g'')\label{a4}
\eeqar
The integration measure $d\mu (g')$ is the Haar
measure on $S^2 =SU(2)/U(1)$ divided by $2j+1$.
In the integration involved, one can extend this to
$G=SU(2)$ since the effect of an $H$-transformation
cancels out in ${\cal D}^{(j)}_{nj} (g')
{\cal D}^{*(j)}_{kj} (g')$. We can then use the 
orthogonality properties of
${\cal D}^{(j)}_{mk} (g')$ to arrive at (\ref{a4}).

Eventhough it is beside our main line of
reasoning, it may be interesting to note that
the start product on $S^2$ can be easily
represented in this framework.
The classical function or symbol corresponding to the
product ${\hat f} {\hat h}$ is
\beqar
(fh)&=& \sum_{mnl} f_{mn} h_{nl}{\cal D}^{*(j)}_{mj} (g)
{\cal D}^{(j)}_{lj} (g) \nonumber\\
&=& \sum_{mnkrl} f_{mn} h_{kl}{\cal D}^{*(j)}_{mj} (g)
{\cal D}^{(j)}_{nr}(g){\cal D}^{*(j)}_{kr}(g)
{\cal D}^{(j)}_{lj} (g) \label{a5}
\eeqar
where we used ${\cal D}^{(j)}_{nr}(g){\cal D}^{*(j)}_{kr}(g)
=\delta_{nk}$. The term $r=j$ in the summation over $r$ gives the
product of the symbols for ${\hat f}$ and ${\hat h}$.
The terms with $r\neq j$ may be rewritten using
\beq
{\cal D}^{(j)}_{n,j-s}(g)=\sqrt{ (2j-s)! \over s! (2j)!}
~~K^s_- \cdot {\cal D}^{(j)}_{nj}(g)\label{a6}
\eeq
We can rewrite (\ref{a5})
\beq
(fh)= \sum_{s=0}^{2j} \left[{ (2j-s)! \over s! (2j)!}\right]
\left[ \sum f_{mn}
{\cal D}^{*(j)}_{mj} (g) K^s_- {\cal D}^{(j)}_{nj}(g)\right]~
\left[ \sum h_{kl}(K^s_-{\cal D}^{(j)}_{kj})^*(g)
{\cal D}^{(j)}_{lj} (g) \right]\label{a7}
\eeq
This gives the star product of two functions on $S^2$.

Functions  and star products on other coset spaces
of unitary groups may be considered in a similar way,
by writing a function as
\beq
f(g,g') = \la g \vert {\hat f}\vert g' \ra =
\sum_{MN} f_{MN} {\cal D}^{*(r)}_{MS} (g)
{\cal D}^{(r)}_{NS} (g') \label{a8}
\eeq
where $r$ denotes a fixed representation and the other
indices are composite indices labelling the states
uniquely. ${\hat f}$ must be invariant under the $H$-subgroup.
As mentioned before, for ${\bf CP}^2$, the representations are sequences
of totally symmetric ones. For the right indices on the
${\cal D}$'s, indicated by $S$ above, we choose the highest weight
states; ${\hat f}$ must have
with overall invariance under $H$.

Derivatives on functions are defined by commutators with the $K$'s
in $\underline{G}- \underline{H}$.
i.e., 
\beq
\partial_i ~f = ~[K_i,~f]\label{a9}
\eeq
The
commutator of two such derivatives is given by
an element of $\underline{H}$, say $[K_i, K_j]= C^\alpha_{ij}K_\alpha$,
$K_\alpha \in \underline{H}$,  and so it
vanishes on a scalar function $f$ (which is invariant under $H$).
\beq
(\partial_i ~\partial_j - \partial_j ~\partial_i )\cdot 
f=0 \label{a10}
\eeq

In describing gravitational fluctuations, we need to go beyond
scalar functions to vectors, tensors, etc. These have a frame
dependence and so they do not commute with elements of
$\underline{H}$. They are specified by the representation
according to which they transform under the $H$-action.
They are thus of the form
\beq
f^{(K)}(g,g') = \sum_{PQ} f_{PQ} {\cal D}^{*(r)}_{P,P'}(g)
{\cal D}^{(r)}_{Q,Q'}(g') \label{a11}
\eeq
where the right indices $P', Q'$ no longer correspond to the same
highest weight value but are such that $f$ transforms as 
some nontrivial representation
$K$ under the action of $H$.
In taking products as in (\ref{a4}), the integration 
measure will be that of $G/H$.
In the 
example of the noncommutative $S^2$, for
$K_3 =-1$,
we have
\beq
f(g,g')= \sum_{mn} f_{mn} {\cal D}^{*(j)}_{mj}(g){\cal D}^{(j)}_{n,j-1}(g')
\label{a12}
\eeq
The corresponding symbol may be written as
\beq
f \sim f^{\beta_1 \beta_2\cdots \beta_{2j}}_{\alpha_1
\alpha_2\cdots \alpha_{2j}} {\bar u}^{\alpha_1}
{\bar u}^{\alpha_2}\cdots {\bar u}^{\alpha_{2j}}
u_{\beta_1}u_{\beta_2}\cdots u_{\beta_{2j-1}}~{\bar u}^{\gamma}\epsilon_{\gamma
\beta_{2j}}~e^{-i\theta /2}
\label{a13}
\eeq
using the parametrization
\beq
g = \left( \matrix{u_1 &u^*_2\cr u_2 & -u^*_1\cr}\right)
~\left(\matrix {e^{i\theta /2}&0\cr 0&e^{-i\theta /2}\cr}\right)
\label{a14}
\eeq
Here $\alpha_i, \beta_i$ take the values $1,2$.
Because of the extra ${\bar u}$ in
(\ref{a13}), the lowest spin value (for the action of $J_a$) is $1$ rather than zero.
$f$ in (\ref{a13}) also has a natural interpretation as rectangular
matrices. (Star products, derivatives and vector bundles have 
also been discussed in
\cite{baletal}. Tensor
analysis on some quantum spaces was considered in \cite{ho}.)

What we have given
above  is a description in terms of local coordinates. For some
of the spaces, one can actually  have a more global description
by considering all the generators of $G$, but with the
specification of some algebraic restrictions on them.
For example, it is well known that in the case of $S^2$, we can
use $SU(2)$ generators $J_i,~i=1,2,3$, with 
$x_i =J_i/ \sqrt{J^2}$ obeying $x_i x_i=1$. For ${\bf CP}^2$, an
analogous description is given by considering eight matrices
$X_A$, $A=1,2,...,8$, with $\sum_A X_A X_A = 2/3$ and $3\sqrt{2}
L^2 - L - \sqrt{2}/3 =0$ where $L=\sum_A X_A \lambda_A$,
$\lambda_A$ being the standard Gell-Mann matrices of $SU(3)$.

Vectors and tensors, as described above, are in the
the analogue of the coordinate basis.
In the following, while using this coordinate basis
for derivatives, etc., 
we shall introduce the gravitational
degrees of freedom in terms of the tangent frame and the group
acting on it. Towards this,  introduce a covariant derivative
\beq
{\cal D}_i = \partial_i ~+~e_i^a (x) ~T^a ~+~ 
\Omega_i^\alpha ~I^\alpha
\label{3}
\eeq
with a gauge field $A_i (x) = e_i^a (x) ~T^a ~+~ \Omega_i^\alpha
~I^\alpha$. Here $T^a$ are the analogues of the $t_i$ but acting
on the tangent frames. $(e_i^a(x), ~\Omega_i^\alpha (x))$ are
functions of
$x_i$ which must also behave as vectors in the coordinate basis.
In the noncommuting case, since
$(e_i^a(x), ~\Omega_i^\alpha (x))$ are not mutually commuting
functions, the gauge group must be a unitary group so that
$[\cD_i,\cD_j]$ is an element of its Lie algebra. In particular,
$(T^a,I^\alpha )$ must belong to the fundamental representation
of this group. $T^a$ being the analogues of
$t_i$, we take this group to be a copy of $U(k)$, which we denote
by
$U_R(k)$ to distinguish it from the $U(k) \subseteq U(N)$
discussed earlier (containing $G$) which will be denoted by
$U_L(k)$ from now on.
$(T^a,I^\alpha )$ then form a basis for $U_R(k)$. At this stage
we are led to $U_L(k)\times U_R(k)$. The coordinates $x_i
\in
\underline{G}-\underline{H}$ and the tangent frame group is
$U_R(k)$.  Since we also have an embedding of $U_L(k)$ in $U(N)$,
$\cD_i$ actually belong to  the algebra of $U(N)\times U(k) =
U(N)\times U_R(k)$. 

The commutator of two covariant derivatives can be expanded as
\beq
[\cD_i ,\cD_j] ~=~ \omega_{ij}~+~{\cal T}_{ij}^a ~T^a~+~ {\cal R}_{ij}^\alpha ~I^\alpha 
\label{4}
\eeq
where $[\del_i,\del_j]=\omega_{ij} =C^\alpha_{ij}K_\alpha$.
${\cal T}_{ij}^a$ is
identified as the torsion tensor and
${\cal R}_{ij}^\alpha$ is the Riemannian curvature. If we take $e_i^a
=\delta_i^a$ and $\Omega_i^\alpha =0$, then $\cD_i 
=\partial_i +T_i$.
Acting on a vector of the form $V= V^kT_k$, we find
\beq
[\cD_i, \cD_j ] \cdot V = [[T_i, T_j] ,T_k] ~V^k
\label{5}
\eeq
where we used the fact that $V^k$ are functions and so
$[[\del_i,\del_j],V^k]=0$. This equation
identifies the curvature as $[T_i,T_j]$, which is not zero in
general. Notice that $e_i^a =\delta_i^a,~ \Omega_i^\alpha =0$
does not correspond to flat space.
The matrices $(e_i^a(x), ~\Omega_i^\alpha (x))$
are the gravitational degrees of freedom on the noncommuting
space. (Usually the gravitational degrees of freedom are
expressed in terms  of deviations from a flat space, but one can
equally well specify them in terms of deviations from any chosen
base space.) In contrast with the commuting limit, the tangent
frame group has to be
$U(k)$ rather than the Poincar\'e group of appropriate dimension.

There are two sets of transformations which are of interest in
terms of their action on $\cD_i$. Transformations of the form
$U(N)\otimes {\bf 1}$ do not act on the tangent frames but
redefine the notion of $\partial_i$.
These can thus be considered as the noncommutative version of
diffeomorphisms. With a chosen set of $\partial_i$,
one can also consider
transformations of the form ${\cal F}\otimes H_R$ where ${\cal F}$'s commute
with $\underline{H}$ and so are functions on the noncommutative
$G/H$-space. These correspond to the local Lorentz
transformations in the usual description of gravity in terms of
frame fields.

\vskip .1in
\noindent{\large\bf 3. Two gauge theories}
\vskip .1in
We must now turn to the choice of an action for the gravitational
degrees of freedom. $(e_i^a, ~\Omega_i^\alpha )$ form the
potential of a
$U(k)$-gauge field. The derivative of a function, $\partial_i f$,
does not necessarily commute with $\underline{H}$, but the
derivatives on any function do commute. So as in the commutative
limit, antisymmetrization of indices lead naturally to
$\underline{H}$-invariant quantities on the noncommutative
$G/H$-space. The only intrinsically defined action then has to be
a Chern-Simons like action or something related to it.

The CS action in $2n+1$ dimensions for gauge fields on a flat
noncommutative space of dimension $2n$ and with one commuting time
coordinate has been given in \cite{poly} as
\beq
\S_{2n+1} = {\lambda \over \mu}\int dt~ \sum_{r=0}^n ~ {(-1)^r
\over 2 r+1} {(n+1)! \over r! (n-r)!} ~\Tr \left( \omega^{n-r}
~\cD^{2r+1}\right)
\label{6}
\eeq
where antisymmetrization of all indices is implicit. $\cD$ is an
antihermitian infinite-dimensional matrix, it may be taken to be
of the form
$\partial_i +A_i$ for some hermitian matrix potential $A_i$.  For
the flat noncommutative space $[\partial_i, \partial_j ]=
\omega_{ij}$ is proportional to the identity. This is not the
case for the general
$G/H$-spaces we are considering, nevertheless $[K_i,K_j]=
C^\alpha_{ij}K_\alpha$
commutes with all functions on the coset space and we can use
essentially the same action $\S$, with the proviso that the number of
$\partial_i$'s has to be even. We take
$\cD$'s to be antihermitian
$M\times M$-matrices which may be considered to be of the form
$\partial_i +A_i$ again, but otherwise the action is unchanged.
Notice that, eventhough $[[\del_i,\del_j], \partial_k] =
C_{ijk}^l\partial_l$ is not necessarily zero, we have
$\ep^{ijkm...} C_{ijk}^l =0$ by the Jacobi identity. Further,
since $A_i$ are covariant vectors in the coordinate basis,
$[\omega_{ij}, A_k] = C_{ijk}^lA_l$ and so, by the previous
identity, $\ep^{ijkm...} [\omega_{ij} , A_k] =0$. Thus
$\ep^{ijkm...}[\omega_{ij} , \cD_k]=0$. The $\cD$'s in the CS
action, by virtue of the antisymmetrization of indices, commute
with
$\omega_{ij}$ and the relative order of $\omega$'s and $\cD$'s in
(\ref{6}) is immaterial. The constant $\mu$ in (\ref{6}) is taken
to be
\beq
\mu = i~{(n+1)~\Tr(\omega^n)\over M} \label{7}
\eeq
The action may be written out in the three-dimensional and
five-dimensional cases, which we use later, as
\beqar
\S_3&=& {\lambda \over \mu} \int dt~\Tr \left( - {2\over 3} \cD^3
+2
\omega
\cD \right)\nonumber\\
\S_5&=& {\lambda \over \mu} \int dt~\Tr \left({3\over 5}\cD^5 - 2
\omega \cD^3 +3
\omega^2 \cD\right)\label{8}
\eeqar

The
$r=0$ term in (\ref{6}) is $ \lambda ~(n+1)\Tr (\omega^n
A_0)/\mu$.  Recall that on the plane, nontrivial $U(1)$ gauge
transformations led to the quantization of the level number
$\lambda$ of the CS term \cite{np}. The elementary nontrivial
transformation is $A_0 \rightarrow A_0 +i~\partial_0 \varphi$,
with $\varphi (t=\infty ) - 
\varphi (t=-\infty ) = 2\pi /M$. With $\mu$ as given in (\ref{7}),
$\Delta \S = 2\pi \lambda$ and the singlevaluedness of $\exp(i\S
)$ gives the quantization of the level number, as in the case of
the flat noncommutative space.

The CS action in (\ref{6}) is very general. $\cD$'s are general
$\underline{U(M)}$ matrices. The only indicator of the space on
which the fields are defined is in fact in the $\omega_{ij}$
appearing in the action. This enters via the definition of the
field strength $F_{ij}= [\cD_i, \cD_j] -\omega_{ij}$. The choice
of different $\omega_{ij}$ and the corresponding definition of
the $\partial_i$'s determine the base space on which the fields
are defined and identifies the particular vacuum state of zero
field strength. Once the $\partial_i$'s are chosen, one can
introduce the splitting $\cD_i =\partial_i +A_i$, identifying the
potential $A_i$. At this stage, there is still no specific choice
of the gauge group made. If we take the matrices $A_i$ to be in
the algebra of $U(N)\times U(k)$, taking $M= Nk$, and approach
the commuting limit, then we get a gauge group $U(k)$.

We now turn to a variant of the CS action which we 
can use in even
dimensions. The variation of the CS action gives
\beq
\delta \S = (n+1) ~\int dt~\Tr \left( \delta \cD ~ F^n \right)
\label{9}
\eeq
We can therefore define an action in $2n$ dimensions by
\beq
\S = \alpha ~\Tr \left( Q ~ F^n \right)
\label{10}
\eeq
where $\alpha$ is a constant and $Q$ is a matrix, $\neq 1$, which
commutes with
$\omega_{ij}$. (A term with $Q=1$ gives a purely ``topological"
action with no contribution to the equations of motion.) This is
equivalent to taking the fields in the CS Lagrangian to be
independent of time and assigning a value $Q$ to the
$A_0$. If the $\cD$'s commute with $Q$, this term is actually
zero, by cyclicity of trace; it is thus the matrix or
noncommutative version of a total derivative. $\cD$'s do not
commute with $Q$ in general, and this theory is a nontrivial
gauge theory in even dimensions. With the insertion of $Q$ in
(\ref{10}), we have gauge invariance  only under transformations
which commute with $Q$. If there are many choices for $Q$, one
can  take a linear combination, interpreting the coefficients as
different coupling constants.  The theory depends on the choice
of 
$Q$, which is arbitrary except that it should commute with
$\omega_{ij}$. We may in fact consider
$Q$ as the quantity which is primarily chosen and then identify
$\omega_{ij}$  in terms of matrices which commute with it.  The
equations of motion for the gauge theory (\ref{10}) is now given
by
\beq
\sum_{r=0}^n~ F^{n-1-r}~[\cD,Q]~F^r =0
\label{11}
\eeq
where the antisymmetrization of indices, as in the actions
(\ref{6}, \ref{10}), is assumed. These equations, along with the
condition $[\omega_{ij}, \cD_k] = C_{ijk}^l \cD_l$, give our
definition of noncommutative gravity.

The gauge theory (\ref{10}) is reminiscent of the gauge theory
approach to constructing the gravitational action in four
dimensions due to Chang, MacDowell and Mansouri (CMM) \cite{CMM}.
In this case, one considers $S^4= SO(5)/SO(4)$ with $SO(4)$
generators
$J^{ab}$ and $S^4$-translations $P^a$ with $[P^a ,P^b ] =\Lambda
J^{ab}$. An explicit realization of these is given by the
$4\times 4$ Dirac matrices, $P^a= i\sqrt{\Lambda}~ \gamma^a$ and
$J^{ab} = -[\gamma^a , \gamma^b]\equiv -\gamma^{ab}$. The
covariant derivatives are
$\cD_\mu =\partial_\mu + e_\mu^a i\sqrt{\Lambda}~\gamma^a
+\Omega_\mu^{ab}(-\gamma^{ab})$. The commutators give
\beqar
[\cD_\mu ,\cD_\nu ] &=& {\cal T}_{\mu\nu}^a ~(i\sqrt{\Lambda}\gamma^a )+
{\cal R}_{\mu\nu}^{ab}~( -\gamma^{ab} )\nonumber\\
{\cal R}_{\mu\nu}^{ab}&=& R_{\mu\nu}^{ab} +\Lambda e_\mu^{[a} e_\nu^{b]}
\nonumber\\
R_{\mu\nu}^{ab}&=& \left(\partial_\mu \Omega_\nu -\partial_\nu \Omega_\mu
+[\Omega_\mu ,\Omega_\nu ] \right)^{ab}
\label{12}
\eeqar
The action is then taken as
\beq
\S = \alpha ~\int  {\cal R}^{ab} {\cal R}^{cd} ~\epsilon_{abcd}
\label{13}
\eeq
Here ${\cal R}^{ab}$ is the curvature two-form with the components
given (\ref{12}). The leading term involving just the spin
connection is a topological invariant and may be dropped as far
as the classical theory is concerned. The remainder gives the
Einstein-Hilbert action with a cosmological constant. If we take
the limit $\Lambda \rightarrow 0$,
$\alpha \rightarrow \infty$ with $\alpha \Lambda$ fixed, we get
the action with zero cosmological constant. This action may be
written, upto an overall constant normalization factor, as
\beq
\S = \alpha \int  \Tr (\gamma^5 \cD^4)
\label{14}
\eeq
This is of the form (\ref{10}) with $Q=\gamma^5$. Notice that
$\gamma^5$ commutes with the subgroup $H= SO(4)$; also we do not
have any
$\omega_{ij}$-type terms since $[\partial_\mu , \partial_\nu ]=0$.
(The observation that the trace with $\gamma^5$ could be used
to obtain the $\epsilon$-tensor in (\ref{13}) was also made
in \cite{gotzes}; I thank M-I Park for bringing this paper to
my attention.)
In the noncommutative case, there is only one trace
which covers the integration over the space as well as trace
over any internal indices. We can thus regard the action
(\ref{10}) as a generalization of  the gauge theory approach to
gravity.

Eventhough the connection with the CMM approach to gravity has
been mentioned, it should be noted that, as in the case of the CS
action, the  action (\ref{10}) is very general, the only
indicator of the base space being the
$\omega_{ij}$ and
$Q$ which commutes with them.

In adapting these actions (\ref{6}, \ref{10}) for  describing
gravitational fluctuations, we can then say that the key step is
the choice of the base space. This can be done by writing
$U(M)$ matrices as $U(N)\times U_R(k) $ matrices and specifying
$H\subset G \subseteq U_L(k) \subseteq U(N)$ and identifying the
derivatives $\partial_i$ with elements complementary to
$\underline{H}$ in
$\underline{G}$. A similar splitting can be made in the other
copy of
$U(k)$, namely $U_R(k)$, with an $H_R \subset U_R(k)$. For the
even dimensional case,
$Q$ is chosen to be an element of $U_R(k)$ which commutes with
$H_R$.

An alternate way of arriving at this action would be as follows.
We choose $Q$ as a $k\times k$-matrix and then identify the
closed subalgebra of $k\times k$-matrices which commute with $Q$.
The algebra $H_R$ is either this algebra or a smaller closed set,
for which we can find a $G$ with $G/H$ being a symmetric space.
We then consider a copy of this structure, namely
$U_L(k)$ and
$H_L \subset G \subseteq U_L(k)$ and identify the derivatives
with the elements of the algebra $\underline{G}$ which are not in
$\underline{H_L}$. This essentially identifies the space on which
the theory is to be defined. We then choose an
$N$-dimensional representation of
$U_L(k)$, for an appropriate $N$, with $M= Nk$ and $\cD$'s in the
algebra of
$U(N)\times U_R(k)$, to obtain the action (\ref{10}). The case
when $H_R$ can be taken to be the full algebra of 
$k\times k$-matrices commuting with $Q$ corresponds to a space 
whose commutative limit is $SU(k)/U(k-1) \sim {\bf CP}^{k-1}$.
This situation is particularly natural because the gravitational 
fields will correspond to  just those in the commuting limit,with
no additional degrees of freedom, except for an overall $U(1)$
field. Notice that the choice of $Q$ is the key, the rest of the
required structure is naturally determined from it.

The basic suggestion of this paper is that CS gravity in odd
dimensions can be described by the CS action in (\ref{6}), with
$\partial_i \in \underline{G}-\underline{H}$, $G\subseteq U_L(k)
\subset U(N)$, with $D_i$ belonging to the algebra of $U(N)\times
U_R(k)$. If $[[\partial_i, \partial_j],\partial_k] =C_{ijk}^l
\partial_l$, then the
$\cD$'s must also obey $ [[\partial_i, \partial_j], \cD_k] =
C_{ijk}^l
\cD_l$. For odd dimensions, there will be a quantization
constraint on the gravitational coupling, since the level number
of the CS action is quantized. In even dimensions, we can define
a gravitational action which is (\ref{10})  with the same
structure for the $\cD$'s as given above for the odd dimensional
case. Since the scale of $Q$ is arbitrary, or since the action is
gauge invariant, with no extra factors like $\Tr (\omega^n A_0)$,
we do not expect quantization of the coefficient. Finally, it is
easy to see that the actions (\ref{6}, \ref{10}) can be written
down directly for the case where the underlying noncommutativity
structure  is that of the Heisenberg algebra. In this case
$\omega_{ij}$ trivially commute with  the $\cD$'s.  Actions
constructed via the star product formalism seem to  involve
determinants and inverses, which are evaluated by expansions via
star products. Our action (\ref{10}) avoids some of this
awkwardness.

In a full theory of gravity, it should be possible to change the
choice of $Q$, $G$, $H$, etc., which would correspond to
transitions between spaces with different topologies in the
commuting limits and even between spaces of different dimensions.
In fact, since $Q$ naturally determines the rest of the
structure, we can say that, if the short distance theory of
gravity involves noncommutative spaces,   then initially we could
have random values for $Q$ following some thermal distribution
and as the universe cooled down,  a particular choice of $Q$,
perhaps randomly as in the case of vacuum orientation for 
spontaneously broken symmetries, was made and this determined the
number of  dimensions of the world. The actions we have given
correspond to expansions around chosen base spaces and describe
the gravitational fields which are small in the sense of not
changing these structures under fluctuations. Further our actions
do not involve the notion of the metric explicitly. When the
coordinates do not commute among themselves, we do not even have
the usual notion of a metric or distance function on the space.
Therefore, a gravitational action which does not use any such
additional structures is more natural, even more so than in the
case of commutative spaces.
\vskip .1in
\noindent{\large\bf 4. Special cases}
\vskip .1in
Consider the case of the noncommutative ${\bf CP}^2$ for which 
$G= SU(3)\subset U(3)$, $H=U(2)$. 
We can write
the covariant derivatives in the form (\ref{3}) with the generators
of $U_R(3)$ taken as
\beqar
I^8 &&\!\!\!\!\!= -{i\over 2\sqrt{3}}\left(\matrix{1&0&0\cr 0&1&0\cr
0&0&-2\cr}\right),\hskip 1in
I^i =-{i\over 2} 
\left(\matrix{\sigma^i&0\cr 0&0\cr}\right)\nonumber\\
T^1 &&\!\!\!\!\!= {1\over \sqrt{2}}
\left( \matrix{ 0&0&1\cr 0&0&0\cr 0&0&0\cr}
\right), \hskip 1in T^2 = {1\over \sqrt{2}}
\left( \matrix{ 0&0&0\cr 0&0&1\cr 0&0&0\cr}
\right)
\label{14a}
\eeqar
where $\sigma^i, ~i=1,2,3$, are the Pauli matrices. 
 $T^{\bar i}$ are the conjugates of the $T^i$. We also have $I^0=
-(i/\sqrt{6})~{\bf 1}$. The matrix $Q$ can be taken to be $I^8$
and we can write the action as
\beq
\S = i \alpha ~ \Tr \left( I^8 F_{\mu\nu} ~F_{\alpha \beta}\right)
\ep^{\mu\nu\alpha\beta}\label{14b}
\eeq
The equations of motion (\ref{11}) simplify as
\beq
\ep^{\mu\nu\alpha\beta} \left\{ ~[\cD_\alpha, I^8] ~F_{\mu\nu}
+F_{\mu\nu} ~[\cD_\alpha, I^8] ~\right\} =0
\label{14c}
\eeq
The action (\ref{14b}) contains a topological term which does not 
contribute to the equations of motion. Therefore, as far as
classical physics is concerned, we can simplify the expression
(\ref{14b}). Separate the generators into $\{ ~I^A ~\}$ which
include $I^i,~ I^8$ and  $I^0$ and $(T^i,~T^{\bar i})$.
$I^A$ commute with $I^8$ and also the products $I^A T^i$,
$I^AT^{\bar i}$ can be expanded in terms of $T^i$ and $T^{\bar
i}$, with no $I^A$'s. We write
\beqar
{\cal D}_\mu &=& \partial_\mu \Omega_\mu 
+E_\mu \equiv D_\mu +E_\mu\nonumber\\
D_\mu &=& \partial_\mu +\Omega_\mu 
=\partial_\mu + \Omega_\mu^A I^A \nonumber\\
E_\mu &=& {\bar e}_\mu^a T^a + e_\mu^a T^{\bar a}\label{14d}
\eeqar
The commutators can now be separated as
\beqar
[D_\mu, D_\nu ] &=& \omega_{\mu\nu} 
~+~ R_{\mu\nu}^A I^A \nonumber\\
~[D_\mu , E_\nu ] - [D_\nu , E_\mu ] 
&=& {\cal T}_{\mu\nu}^i T^{\bar i} 
+ {\bar{\cal T}_{\mu\nu}^i} T^i \nonumber\\ 
~[E_\mu ,E_\nu ] &=&  \Lambda_{\mu\nu}^A I^A \label{14e}
\eeqar
In the action, since $D_\mu$ commutes with $I^8$, one has the
identity
$\ep^{\mu\nu\alpha\beta}\Tr (I^8 [D_\mu, D_\nu] [D_\alpha,
D_\beta ])=0$, which leads to $\ep^{\mu\nu\alpha\beta}\Tr(I^8
R_{\mu\nu}R_{\alpha\beta})=0$. The term quadratic in the torsion
can be simplified as
\beqar
\ep^{\mu\nu\alpha\beta}\Tr \left( I^8 ({\cal T}_{\mu\nu}^i T^{\bar i} 
+ {\bar{\cal T}_{\mu\nu}^i} T^i)({\cal T}_{\alpha\beta}^j T^{\bar j} 
+ {\bar{\cal T}_{\alpha\beta}^j} T^j)\right) &=&
\ep^{\mu\nu\alpha\beta}{i\over \sqrt{3}} \Tr ([D_\mu , E_\nu ][D_\alpha , E_\beta ])\nonumber\\
&=&-{i\over 2\sqrt{3}}\ep^{\mu\nu\alpha\beta}\Tr [D_\mu ,D_\nu ] [E_\alpha ,E_\beta ]\nonumber\\
&=&-{i\over 2\sqrt{3}} \ep^{\mu\nu\alpha\beta}\Tr ( R_{\mu\nu}\Lambda_{\alpha\beta})
\label{14f}
\eeqar
where we have used the Jacobi identity and the fact that
$\ep^{\mu\nu\alpha\beta}[\omega_{\mu\nu}, E_\alpha ] =0$ by
virtue of the assigned transformation law for $E_\alpha$. Using
these results, the action (\ref{14b}) can be written, upto total
derivatives, as
\beqar
\S&=& {i\alpha\over 4} \Tr \left( J [D_\mu ,D_\nu ] [E_\alpha ,E_\beta ] ~-~ J \omega_{\mu\nu}
[E_\alpha ,E_\beta ] ~+~ I^8 [E_\mu ,E_\nu ][E_\alpha ,E_\beta ]\right)\ep^{\mu\nu\alpha\beta}
\nonumber\\
J&=& 2 I^8 - {i\over 2\sqrt{3}} {\bf 1}\label{14g}
\eeqar 

A simple solution of these equations is given by a pure gauge
form
\beq
\cD_\mu = M^{-1} (\partial_\mu ) M = \partial_\mu +
M^{-1}[\partial_\mu ,M]
\label{14g2}
\eeq
where $M$ is a matrix which is valued
in $U(3)_R$ and is a function on the noncommutative ${\bf
CP}^2$. In other words, we can write
\beq
M= \exp\left( \theta_i T^i +{\bar \theta}_i T^{\bar i}
+\varphi_A I^A\right) \label{14h}
\eeq
where $\theta_i , {\bar \theta}_i, \varphi_A$ are $(N\times
N)$-matrices which commute with $\omega_{\mu\nu}\in \underline{H}$. When
$\theta_i ={\bar \theta}_i =0$, this correponds to a local
Lorentz-type transformation; the relevant degrees of freedom are
thus in the $\theta_i , {\bar \theta}_i$. Expanding
\beq
M^{-1}[\partial_\mu ,M]= {\bar e}_\mu^a T^a + e_\mu^a T^{\bar a}
+\Omega_\mu^A I^A
\label{14i}
\eeq
we can see that $\half ({\bar e}_\mu^a e_\nu^a + e_\nu^a {\bar
e}_\mu^a )$ defines the ${\bf CP}^2$-metric tensor in the
commuting limit. The configuration (\ref{14g2}, \ref{14h})
thus corresponds to the noncommutative ${\bf CP}^2$ again.

We have formulated the gravitational field
equations in (\ref{14b}) parametrizing
the fields in terms of deviations from a noncommutative ${\bf
CP}^2$. The result given above is then equivalent to saying that
the
noncommutative
${\bf CP}^2$ itself is a solution of these gravitational
equations; it is in a sense the analogue of the `flat space'.
It is well known that ${\bf CP}^2$ is a gravitational instanton
in the standard commutative gravity with a cosmological term.
What we have shown is that a noncommutative version of this
statement holds as well.  
It would be interesting to
explore other solutions of this equation;  so far, we have not
been able to find any other particularly interesting solutions.

Another interesting example is given by a noncommutative
four-sphere defined by taking
$H =SO(4)$, $G= SO(5) \subset U_L(4)$. $U_R(k)$ is $U(4)$, the
algebra of which is given by the four-dimensional Dirac
$\gamma$-matrices. In this case, $\omega_{ij}$ are generators of
$O(4) \subset U_L(4)$. As a basis for $U_R(4)$ we can take 
\beqar
T^5 =i\gamma^5, \hskip .5in &&\hskip .5in T^0 =i {\bf 1}
\nonumber\\
T^a=i\gamma^a ,\hskip .5in &&\hskip .5in {\tilde T}^a =\gamma^a \gamma^5\nonumber\\
J^{ab}&=& -{1\over 8}[\gamma^a ,\gamma^b] 
\label{18}
\eeqar
The action (\ref{10}) is given by
\beq
\S = \alpha ~ \Tr \left( \gamma^5 F_{\mu\nu}~F_{\alpha\beta}
\right)~\ep^{\mu\nu\alpha\beta}
\label{15}
\eeq
$\partial_\mu$ are $4N\times 4N$-matrices of the form
$\partial_\mu \otimes {\bf 1}_{4\times 4}$. We also have
\beq
[\omega_{\mu\nu}, \partial_\alpha ] =4 \left(\delta_{\mu\alpha}
\partial_\nu - \delta_{\nu\alpha} \partial_\mu \right)
\label{16}
\eeq
The equations of motion for (\ref{15}) are
\beq
\ep^{\mu\nu\alpha\beta} \left\{ ~[\cD_\alpha, \gamma^5] 
~F_{\mu\nu}
+F_{\mu\nu} ~[\cD_\alpha,\gamma^5] ~\right\} =0
\label{17}
\eeq

For ${\bf CP}^2$, the relevant gauge field is a $U(3)$ gauge
field, which has no  additional gravitational degrees of freedom
compared to the commuting limit,  except for an overall $U(1)$
field. This is because $G$ is already a special unitary group, 
namely $SU(3)$. The situation with
$S^4$ is not so nice as the case of
${\bf CP}^2$. We can consider $S^4$ as $SO(5)/SO(4)$, then
$G=SO(5)$ has to be embedded in $U(4)$ and a gauge theory of
$U(4)$ constructed. There are therefore additional degrees of
freedom. The desire to eliminate them has led to recent attempts
at the so-called teleparallelism theories, where the extra
degrees of freedom as set to zero \cite{moffcham, nishino}. An
analogous restriction can be made in our case, by only including
the $e_\mu^a, ~\Omega_\mu^{ab}$ components for the gauge field.
There will still be components of the curvature corresponding to
$T^5, {\tilde T}^a$ and ${\bf 1}$  directions, but these will
vanish as the commutative limit is approached. The action
(\ref{14b}) can still be used, but the equations of motion are
more restricted than (\ref{17}), because there are less number
of fields to vary. Restricting to
$e_\mu^a, ~\Omega_\mu^{ab}$, the various components of the
curvature can be worked out as
\beqar
{\cal T}_{\mu\nu}^a &=& \partial_\mu e_\nu^a - 
\partial_\nu e_\mu^a +{1\over 2}\left(
e_\mu^b \Omega_\nu^{ab} + 
\Omega_\nu^{ab}e_\mu^b - e_\nu^b \Omega_\mu^{ab} +
\Omega_\mu^{ab}e_\nu^b
\right)\nonumber\\
{\tilde {\cal T}_{\mu\nu}^d}
&=& {1\over 4}\ep^{abcd}\left( [e_\mu^a ,
\Omega_\nu^{bc}] -[e_\nu^a, \Omega_\mu^{bc}]\right)\nonumber\\
R_{\mu\nu}^0 
&=& -i[ e_\mu^a, e_\nu^a]+{i\over 8}
[\Omega_\mu^{ab},\Omega_\nu^{ab}]\nonumber\\ R_{\mu\nu}^5 &=&
-{i\over 16}\ep^{abcd}[\Omega_\mu^{ab}
,\Omega_\nu^{cd}]\nonumber\\ R_{\mu\nu}^{ab} &=& \partial_\mu
\Omega_\nu^{ab} - 
\partial_\nu \Omega_\mu^{ab} -{1\over 2}\left(
\Omega_\mu^{ac}\Omega_\nu^{cb} - 
\Omega_\nu^{ac}\Omega_\mu^{cb} -\Omega_\mu^{bc}\Omega_\nu^{ca}
+\Omega_\nu^{bc}\Omega_\mu^{ca}\right)\nonumber\\
{\cal R}_{\mu\nu}^{ab}&=& R_{\mu\nu}^{ab} 
+4\left( e_\mu^a e_\nu^b - e_\nu^a e_\mu^b
+ e_\nu^b e_\mu^a - e_\mu^b e_\nu^a \right)
\label{19}
\eeqar
As in the case of ${\bf CP}^2$ some of the terms in the action
vanish by cyclicity of trace. The action (\ref{15}) can then be
simplified as
\beqar
\S &=& 8~\alpha~ \Tr \left\{ \ep^{abcd} \left[ e_\mu^a e_\nu^b
R_{\alpha\beta}^{cd}  +8~ e_\mu^a e_\nu^b e_\alpha^c e_\beta^d
\right] +{i\over 4} {\cal T}_{\mu\nu}^a {\tilde{\cal
T}_{\alpha\beta}^a} +{i\over 4} [e_\mu^a, e_\nu^a ]
R_{\alpha\beta}^5 \right\} \ep^{\mu\nu\alpha\beta}
\nonumber\\
\label{20}
\eeqar Eventhough, the reduction of fields can be implemented as
above,  from the gauge theory point of view, it is more natural
to  keep all the additional  fields corresponding to the 
$T^5, {\tilde T}^a$ and ${\bf 1}$ directions. Perhaps a
different approach might be to consider ${\bf CP}^3$ which
will be a solution in the analogous
six-dimensional theory and
which will not need additional degrees of freedom since ${\bf
CP}^3 = SU(4)/U(3)$. 
${\bf CP}^3$ can also be considered as an $S^2$ bundle over
$S^4$, this is the Penrose projective twistor space for $S^4$.
By some natural way of projecting out the $S^2$,
this might lead to a better way to formulate
$S^4$ \cite{ram}. 

\vskip .1in
I thank A. Polychronakos for comments. This work was supported
in part by the National Science Foundation grant number
PHY-0070883 and a PSC-CUNY-32 award.

\end{document}